\documentclass[12pt]{article}
\usepackage[portrait]{geometry}
\usepackage{amsmath}

\input{tcilatex}
\begin{document}

\title{On the solution of the variational optimisation in the rational
inattention framework}
\author{Nigar Hashimzade\thanks{%
Correspondence to: nigar.hashimzade@durham.ac.uk. I am grateful to Parantap
Basu for bringing this problem to my attention. All errors are mine.} \\
Durham University, Durham, United Kingdom}
\maketitle

\begin{abstract}
I analyse the solution method for the variational optimisation problem in
the rational inattention framework proposed by Christopher A. Sims. The
solution, in general, does not exist, although it may exist in exceptional
cases. I show that the solution does not exist for the quadratic and the
logarithmic objective functions analysed by Sims (2003, 2006). For a
linear-quadratic objective function a solution can be constructed under
restrictions on all but one of its parameters. This approach is, therefore,
unlikely to be applicable to a wider set of economic models.
\end{abstract}

\section{Introduction}

In two prominent papers (Sims 2003, 2006) Christopher A. Sims proposed to
model decision under uncertainty as the optimal choice of the joint
distribution of action $Y$ and external state $X$, under the constraint on
the flow of information. It is assumed that the marginal distribution of $X$
is known, and the information flow is quantified as the mutual information
of $X$ and $Y$, $I\left( X,Y\right) =H\left( X\right) +H\left( Y\right)
-H\left( X,Y\right) =H\left( Y\right) -H\left( \left. Y\right \vert X\right) 
$, where for a random variable $W$ with distribution $p$, $H\left( W\right)
\equiv -E\left[ \log _{2}p\right] $. This approach to optimisation under
uncertainty belongs to a more general concept of \textit{rational inattention%
}\ introduced by Sims, which within the last fifteen years has developed
into a large literature, with applications to consumption, price and wage
setting, and portfolio choice (Wiederholt, 2017).

Examples in Sims (2003, 2006) are maximisation of expected utility or
minimisation of expected loss, with continuous distribution functions. The
objective and the constraint are, therefore, definite integrals of unknown
functions, and the optimisation problem is solved by finding an extremum of
a functional. While in several follow-up applications the optimisation is
carried out numerically, these two papers present analytical
characterisation of the solution for several special cases. However, the
analysis appears to have a fundamental flaw. Below, I outline the framework
proposed by Sims and focus on two examples, a quadratic loss function (Sims
2003) and a two-period model of consumption and savings with logarithmic
utility (Sims 2006).\footnote{%
One of the working paper version of Sims (2006) is Sims (2005). The latter
provides some details of analytical derivations of the results presented in
the former.} The aim of my paper is twofold. First, it shows how the correct
characterization of the solution can be obtained, using these two examples.
Second, it demonstrates the restrictiveness of this framework, which
suggests that it is unlikely to apply to a wider set of objective functions
and distributions arising in economic models.

\section{`Rational inattention' as constrained variational optimisation}

The rational inattention models are built on the assumption that an economic
agent has a limited capacity for processing information when making a
decision. An agent chooses an action taking into account an external state.
The state cannot be perfectly observed, and both the action and the state
are assumed to be random variables. The agent knows the distribution of the
state which is fixed exogenously. The objective of the agent is to maximise
some criterion function, $CF$, such as the expected utility or negative of
the expected loss. Let $Y\in \mathcal{Y}$ be an action in the action space $%
\mathcal{Y}$ and let $X\in \mathcal{X}$ be a state with distribution $%
p\left( x\right) $ defined over space $\mathcal{X}$. Let $f\left( x,y\right) 
$ describe the joint distribution of $X$ and $Y$. The assumed limit on the
agent's capacity to process information is modelled as the constraint on the
mutual information between $X$ and $Y$. Thus, the agent solves%
\begin{equation*}
\max CF=E\left[ U\left( X,Y\right) \right] s.t.I\left( X,Y\right) \equiv E%
\left[ \log _{2}\frac{f\left( x,y\right) }{p\left( x\right) g\left( y\right) 
}\right] \leq \kappa .
\end{equation*}%
where $p\left( x\right) =\int \nolimits_{\mathcal{Y}}dy$ $f\left( x,y\right) 
$ and $g\left( y\right) =\int \nolimits_{\mathcal{X}}dx$ $f\left( x,y\right) 
$ are the marginal distribution. Sims (2003, 2006) suggested to use the
joint distribution as the instrument of optimisation. Since $p\left(
x\right) $ is fixed, this is equivalent to choosing the distribution of $Y$
conditional on $X$. When $X$ and $Y$ are continuous random variables, the
agent's problem is%
\begin{equation}
\max_{q\left( \left. y\right \vert x\right) }\iint_{\mathcal{X\times Y}}dx%
\text{ }dy\text{ }q\left( \left. y\right \vert x\right) p\left( x\right)
U\left( x,y\right) \text{ s.t. }\iint_{\mathcal{X\times Y}}dx\text{ }dy\text{
}q\left( \left. y\right \vert x\right) p\left( x\right) \log _{2}\frac{%
q\left( \left. y\right \vert x\right) }{g\left( y\right) }\leq \kappa
\label{constrmax}
\end{equation}%
where $q\left( \left. y\right \vert x\right) =\frac{f\left( x,y\right) }{%
p\left( x\right) }$ is the conditional distribution of action choice. This
is a constrained optimisation problem of the calculus of variations (see,
for example, Smirnov et al. 1933), since the unknown is a function, and the
objective and the constraint are functionals. The problem in (\ref{constrmax}%
) is equivalent to the maximisation of a Lagrangean,%
\begin{eqnarray}
\mathcal{L} &\mathcal{=}&\iint_{\mathcal{X\times Y}}dx\text{ }dy\text{ }%
q\left( \left. y\right \vert x\right) p\left( x\right) U\left( x,y\right)
\label{main} \\
&&+\widetilde{\lambda }\left[ \kappa -\iint_{\mathcal{X\times Y}}dx\text{ }dy%
\text{ }q\left( \left. y\right \vert x\right) p\left( x\right) \log _{2}%
\frac{q\left( \left. y\right \vert x\right) }{\int \nolimits_{\mathcal{X}%
}dx^{\prime }\text{ }q\left( \left. y\right \vert x^{\prime }\right) p\left(
x^{\prime }\right) }\right] ,  \notag
\end{eqnarray}%
where $\widetilde{\lambda }\geq 0$ is the Lagrange multiplier, such that $%
\widetilde{\lambda }>0$ when the constraint is binding (holds with equality)
and $\widetilde{\lambda }=0$ otherwise. In addition, one needs to specify
some boundary conditions for $q\left( \left. y\right \vert x\right) $. The
natural boundary condition in this setting is the normalisation, 
\begin{equation}
\dint \limits_{\left. \mathcal{Y}\right \vert x}dy\text{ }q\left( \left.
y\right \vert x\right) =1,\forall x\in \mathcal{X}\text{.}  \label{bc1}
\end{equation}

It is known from the calculus of variations that the necessary condition for
an extremum of functional,%
\begin{equation}
\mathcal{F=}\int \limits_{\mathcal{X}}dx\text{ }F\left( x,y\left( x\right)
,y^{\prime }\left( x\right) \right)  \label{f1}
\end{equation}%
of function $y\left( x\right) $, with boundary condition $\left. y\left(
x\right) \right \vert _{\left( x\right) \in \partial \left( \mathcal{X}%
\right) }=y_{0}\left( x\right) $, is given by $\delta \mathcal{F}=0$,
leading to an Euler equation, 
\begin{equation}
\frac{\partial F}{\partial y}-\frac{\partial }{\partial x}\frac{\partial F}{%
\partial y^{\prime }}=0,  \label{ee1}
\end{equation}%
which, in general, can be rewritten as an ordinary differential equation of
second order with respect to $x$. The general solution is a family of
curves, and a particular solution is found from the boundary conditions.
Similarly, the necessary condition $\delta \mathcal{F}=0$ for the extremum
of functional%
\begin{equation}
\mathcal{F=}\iint_{\mathcal{X\times Y}}dx\text{ }dy\text{ }F\left(
x,y,z\left( x,y\right) ,z_{x},z_{y}\right)  \label{f2}
\end{equation}%
of function $z\left( x,y\right) $ of two variables, $x$ and $y$, with
boundary condition $\left. z\left( x,y\right) \right \vert _{\left(
x,y\right) \in \partial \left( \mathcal{X\times Y}\right) }=z_{0}\left(
x,y\right) $, leads to the Euler equation given by 
\begin{equation}
\frac{\partial F}{\partial z}-\frac{\partial }{\partial x}\frac{\partial F}{%
\partial z_{x}}-\frac{\partial }{\partial y}\frac{\partial F}{\partial z_{y}}%
=0,  \label{ee2}
\end{equation}%
which, in general, is equivalent to a partial differential equation of
second order. The general solution is a family of surfaces, and a particular
solution is found from the boundary conditions. For a constrained
optimisation the objective functional includes a term associated with the
constraint with the Lagrange multiplier, and the corresponding first-order
condition is known as the Euler-Lagrange equation.

When the objective function does not contain the derivatives of the unknown
function, the necessary condition for the extremum, $\frac{\partial F}{%
\partial y}=0$ for $y\left( x\right) $ in (\ref{f1}), or $\frac{\partial F}{%
\partial z}=0$ in (\ref{f2}), is not a differential equation. The extremum
in this case is described by $y=\varphi \left( x\right) $ (or, respectively,
by $z=\varphi \left( x,y\right) $), and, in general, the solution does not
exist, although the problem may have a solution in exceptional cases
(Smirnov et al., 1933, p. 14). In other words, an extremum that satisfies
the given boundary conditions may only exist for some exceptional boundary
conditions.

One can see immediately that functional $\mathcal{L}$ in (\ref{main}) does
not contain the derivatives of the unknown function. Therefore, the
Euler-Lagrange equation for this optimisation problem is not a differential
equation, and the solution does not, in general exist, -- in a sense that
function $q\left( \left. x\right \vert y\right) =\varphi \left( x,y\right) $
that maximises $\mathcal{L}$ in (\ref{main}) may not satisfy condition (\ref%
{bc1}).

Suppose, however, that a solution exists for some exceptional case. Then it
must satisfy the Euler-Lagrange equation, which for (\ref{main}) can be shown%
\footnote{%
See Appendix for details.} to have the form 
\begin{equation*}
q\left( \left. x\right \vert y\right) =g\left( y\right) \exp \frac{U\left(
x,y\right) }{\lambda }
\end{equation*}%
with boundary condition (\ref{bc1}), or, equivalently,%
\begin{equation}
h\left( \left. x\right \vert y\right) =p\left( x\right) \exp \frac{U\left(
x,y\right) }{\lambda }  \label{soln}
\end{equation}%
with boundary condition%
\begin{equation}
\dint \limits_{\left. \mathcal{X}\right \vert y}dx\text{ }h\left( \left.
x\right \vert y\right) =1,\forall y\in \mathcal{Y}  \label{bc}
\end{equation}%
where $\lambda \equiv \frac{\widetilde{\lambda }}{\ln 2}$, and natural
logarithm is introduced for convenience in further derivations.

The potential solution is now analysed for two examples of $U\left(
x,y\right) $ presented in Sims (2003, 2006).

\section{Linear-quadratic loss function}

Consider the problem of minimisation of the expected value of a
linear-quadratic loss function\footnote{%
This example can also be interpreted as maximisation of the expected value
of a linear-quadratic utility in a two-period model of consumption and
saving, allowing for negative consumption and wealth; see Sims (2005).},%
\begin{equation*}
U\left( X,Y\right) =-\theta ^{2}Y^{2}+2\varphi YX-X^{2}+2bX+2cY,\text{ }%
\mathcal{X\times Y=R\times R}.
\end{equation*}%
This is a generalisation of the quadratic loss function ($\varphi =\theta =1$%
, $b=c=0$) considered in Sims (2003), where it is stated that `\textit{when
the }$X$\textit{\ distribution is Gaussian, it is not too hard to show that
the optimal form for }$q$\textit{\ is also Gaussian, so that }$Y$\textit{\
and }$X$\textit{\  \ end up jointly normaly distributed}' (p. 670). As I show
below, Gaussian $q$ as a solution of (\ref{main}) given Gaussian $p$ only
exists and satisfies the properties of a distribution function under certain
restrictions on all but one of the loss function parameters.

Let $X\sim N\left( \mu _{x},\sigma _{x}^{2}\right) $. With $N\left( \mu
_{\left. x\right \vert y},\sigma _{\left. x\right \vert y}^{2}\right) $ as a
guess for $h\left( \left. x\right \vert y\right) $, we have 
\begin{equation*}
I\left( X,Y\right) =-\frac{1}{2}\log _{2}\left( \frac{\sigma _{\left.
x\right \vert y}^{2}}{\sigma _{x}^{2}}\right) =\frac{1}{2}\log _{2}\left(
1-\rho ^{2}\right) .
\end{equation*}%
and, setting $I\left( X,Y\right) =\kappa $ gives 
\begin{equation}
\rho ^{2}=1-2^{-2\kappa }.  \label{rho2}
\end{equation}%
Next, using the properties of the conditional and marginal densities of the
bivariate Gaussian distribution\footnote{%
For the conditional distribution the mean and the variance are given by $\mu
_{\left. x\right \vert y}=\mu _{x}+\rho \frac{\sigma _{x}}{\sigma _{y}}%
\left( y-\mu _{y}\right) $ and $\sigma _{\left. x\right \vert y}^{2}=\sigma
_{x}^{2}\left( 1-\rho ^{2}\right) $.} we obtain from (\ref{soln}) the
expression for the Lagrange multiplier,%
\begin{equation}
\widetilde{\lambda }=\frac{2\ln 2}{2^{2\kappa }-1}\sigma _{x}^{2},
\label{lambda}
\end{equation}%
and the following set of relationships among the model parameters (see
Appendix for details):

\begin{eqnarray}
\theta ^{2} &=&\frac{\sigma _{x}^{2}}{\sigma _{y}^{2}},  \label{Y2} \\
\varphi &=&\frac{1}{\rho }\frac{\sigma _{x}}{\sigma _{y}},  \label{XY} \\
b &=&\left( \mu _{x}-\varphi \mu _{y}\right) ,  \label{X} \\
c &=&-\varphi \left( \mu _{x}-\rho ^{2}\varphi \mu _{y}\right) ,  \label{Y}
\end{eqnarray}%
where $\mu _{y}$ is determined from%
\begin{equation}
\frac{1}{2}\ln \frac{1}{1-\rho ^{2}}=\mu _{x}^{2}-2\varphi \mu _{x}\mu _{y}+%
\frac{\sigma _{x}^{2}}{\sigma _{y}^{2}}\mu _{y}^{2}.  \label{muY}
\end{equation}

Equations (\ref{rho2}) and (\ref{Y2})-(\ref{Y}) effectively restrict three
out of four parameters of the loss function, given $\kappa $ and $\left( \mu
_{x},\sigma _{x}^{2}\right) $, for the optimisation problem to have
conditional Gaussian distribution as a solution. Suppose, we fix $\theta $;
this, along with (\ref{rho2}), determines $\varphi $ in (\ref{XY}), and with 
$\mu _{y}$ calculated from (\ref{muY}), determines $b$ and $c$ by (\ref{X})
and (\ref{Y}). One can see that restrictions $\theta =\varphi =1$ and $b=c=0$
cannot hold simultaneously, and so the solution for $q$ in the case of
quadratic loss function $-\left( Y-X\right) ^{2}$ analysed in Sims (2003)
does not exist.

For $\mu _{x}=0$ and $\theta =1$ we have $\mu _{y}=\sqrt{\kappa \ln 2}$ and 
\begin{equation*}
\varphi =\frac{1}{\sqrt{1-2^{-2\kappa }}},\text{ }b=-\sqrt{\frac{\kappa \ln 2%
}{1-2^{-2\kappa }}},\text{ }c=\sqrt{\kappa \ln 2}.
\end{equation*}%
In this case the optimal $q\left( \left. y\right \vert x\right) $ is
Gaussian with%
\begin{eqnarray*}
\mu _{\left. y\right \vert x} &=&\mu _{y}+\sqrt{1-2^{-2\kappa }}x, \\
\sigma _{\left. y\right \vert x}^{2} &=&2^{-2\kappa }\sigma _{x}^{2},
\end{eqnarray*}%
but this solution only exists for 
\begin{equation*}
U\left( X,Y\right) =-Y^{2}+2\frac{1}{\sqrt{1-2^{-2\kappa }}}YX-X^{2}-2\sqrt{%
\frac{\kappa \ln 2}{1-2^{-2\kappa }}}X+2\sqrt{\kappa \ln 2}Y.
\end{equation*}

\section{Logarithmic consumption-savings model}

This example is different in one important way which highlights how
restrictive the variational approach is in the rational inattention
framework. In the previous example the distributions of the state and action
variables allow, in principle, for an unbounded support, and so a solution
could be constructed for a suitable, albeit restricted, choice of the model
parameters. When the nature of economic variables dictates the bounds on the
support of the distribution (for example, non-negativity), the solution may
not exist for any configuration of the remaining model parameters, -- the
existence of bounds, in effect, poses additional restrictions that cannot be
met simultaneously.

The following example of a two-period consumption-savings model with
logarithmic utility was analysed in Sims (2006).\footnote{%
In Sims (2005, 2006) $\beta =1$ and the notations correspond to $w=x$ and $%
c=y$.} An individual with random endowment $X>0$ chooses how to allocate $X$
between consumption, $Y\leq X$, in the first period, and savings, $X-Y$, to
be consumed in the second period. The objective is to maximise the expected
utility function, $E\left[ U\left( X,Y\right) \right] $, where $U\left(
X,Y\right) =\ln Y+\beta \ln \left( X-Y\right) $. The distribution of $X$ is
given by $p\left( x\right) $, and the individual chooses $q\left( \left.
y\right \vert x\right) $ under the constraint on the information flow.

A potential solution for $q\left( \left. y\right \vert x\right) $, if it
exists, must be consistent with (\ref{soln}):%
\begin{equation*}
h\left( \left. x\right \vert y\right) =p\left( x\right) \exp \frac{U\left(
x,y\right) }{\lambda }=p\left( x\right) y^{\alpha }\left( x-y\right) ^{\beta
\alpha },\text{ }0<y<x<\infty \text{.}
\end{equation*}%
This can be rewritten as%
\begin{equation*}
h\left( \left. x\right \vert y\right) =p\left( x\right) \left( \frac{y}{x}%
\right) ^{\alpha }\left( 1-\frac{y}{x}\right) ^{\beta \alpha }x^{\left(
1+\beta \right) \alpha }
\end{equation*}%
Because the support of the distribution is bounded, in order to satisfy (\ref%
{bc}) it must be the case that%
\begin{equation}
p\left( x\right) =\frac{x^{-\left( 1+\beta \right) \alpha -1}}{B\left(
\alpha ,\beta \alpha +1\right) },\text{ }x\in \mathcal{X}.  \label{power}
\end{equation}%
This can be verified directly:%
\begin{eqnarray*}
&&\int_{y}^{\infty }dx\text{ }p\left( x\right) \left( \frac{y}{x}\right)
^{\alpha }\left( 1-\frac{y}{x}\right) ^{\beta \alpha }x^{\left( 1+\beta
\right) \alpha }=\frac{1}{B\left( \alpha ,\beta \alpha +1\right) }%
\int_{y}^{\infty }\frac{dx}{x}\left( \frac{y}{x}\right) ^{\alpha }\left( 1-%
\frac{y}{x}\right) ^{\beta \alpha } \\
&=&\frac{1}{B\left( \alpha ,\beta \alpha +1\right) }\int_{0}^{1}dz\text{ }%
z^{\alpha -1}\left( 1-z\right) ^{\beta \alpha }=1
\end{eqnarray*}%
Thus,%
\begin{equation*}
h\left( \left. x\right \vert y\right) =\frac{x^{-\left( 1+\beta \right)
\alpha -1}y^{\alpha }\left( x-y\right) ^{\beta \alpha }}{B\left( \alpha
,\beta \alpha +1\right) }.
\end{equation*}%
This formally resembles the expression obtained by Sims (2006)\footnote{%
Sims (2005) derives the expression for the conditional density which
contains a Lagrange multiplier on the marginal density constraint. This
appears to be incorrect; see Appendix. Furthermore, the integrand in
equation (7) in Sims (2006) (the same in Sims 2005) is not proportional to
the density of $F\left( 2\alpha +2,2\alpha \right) $ distribution, --
contrary to what is stated in the paper (p. 161). For this to be the case
the term in parentheses in the integrand should be $\left( v+\frac{\alpha }{%
1+\alpha }\right) $, rather than $\left( v+1\right) $.} with $\beta =1$. The
conditional mean of $X$ exists for $\alpha >1$ (that is, for $\lambda <1$,
so $\widetilde{\lambda }<\ln 2$) and is given by%
\begin{eqnarray*}
E\left[ \left. X\right \vert Y=y\right] &=&\int_{y}^{\infty }dx\text{ }%
xh\left( \left. x\right \vert y\right) =\frac{1}{B\left( \alpha ,\beta
\alpha +1\right) }\int_{y}^{\infty }dx\text{ }\left( \frac{y}{x}\right)
^{\alpha }\left( 1-\frac{y}{x}\right) ^{\beta \alpha \varphi } \\
&=&\frac{y}{B\left( \alpha ,\beta \alpha +1\right) }\int_{0}^{1}dz\text{ }%
\left( \frac{y}{x}\right) ^{\alpha -1}\left( 1-\frac{y}{x}\right) ^{\beta
\alpha } \\
&=&\frac{y}{B\left( \alpha ,\beta \alpha +1\right) }B\left( \alpha -1,\beta
\alpha +1\right) =\left( 1+\beta \right) y\frac{\alpha }{\alpha -1},
\end{eqnarray*}%
Thus, $\frac{E\left[ \left. X\right \vert Y=y\right] }{y}=\left( 1+\beta
\right) \frac{\alpha }{\alpha -1}>1+\beta $, whereas the certainty solution
is $x/y=1+\beta $, -- consistent with the argument that the rational
inattention solution is closer to the certainty solution, the lower is the
shadow price of the information constraint.

As shown above, this solution for $h\left( \left. x\right \vert y\right) $
exists if $p\left( x\right) $ is a power law distribution (\ref{power}) with
support $\mathcal{X}=\left[ x_{0},\infty \right) $ for some $x_{0}>0$. The
normalisation condition, 
\begin{equation*}
1=\int \limits_{\mathcal{X}}dx\text{ }p\left( x\right) =\int
\nolimits_{x_{0}}^{\infty }dx\text{ }\frac{x^{-\left( 1+\beta \right) \alpha
-1}}{B\left( \alpha ,\beta \alpha +1\right) }=\frac{1}{B\left( \alpha ,\beta
\alpha +1\right) }\frac{x_{0}^{-\left( 1+\beta \right) \alpha }}{\left(
1+\beta \right) \alpha }
\end{equation*}%
determines the Lagrange multiplier, $\widetilde{\lambda }=\frac{\ln 2}{%
\alpha }$, implicitly as a function of the model parameters:%
\begin{equation*}
\frac{\alpha x_{0}^{\left( 1+\beta \right) \alpha }}{B\left( \alpha ,\beta
\alpha +1\right) }=\frac{1}{1+\beta }.
\end{equation*}%
However, it is impossible to construct the solution for $q\left( \left.
y\right \vert x\right) $ that satisfies boundary condition (\ref{bc1}).
Formally,%
\begin{equation*}
q\left( \left. y\right \vert x\right) =\frac{h\left( \left. x\right \vert
y\right) g\left( y\right) }{p\left( x\right) }=g\left( y\right) y^{\alpha
}\left( x-y\right) ^{\beta \alpha },\text{ }0\leq x_{0}<y<x<\infty ,
\end{equation*}%
and%
\begin{equation*}
\frac{d}{dx}\int \limits_{\mathcal{Y}}dy\text{ }q\left( \left. y\right \vert
x\right) =\frac{d}{dx}\int \nolimits_{x_{0}}^{x}dy\text{ }g\left( y\right)
y^{\alpha }\left( x-y\right) ^{\beta \alpha }=\beta \alpha \int
\nolimits_{x_{0}}^{x}dy\text{ }g\left( y\right) y^{\alpha }\left( x-y\right)
^{\beta \alpha -1}>0
\end{equation*}%
since the integrand is non-negative on $\left[ x_{0},x\right] $ and is
strictly positive at least on some subinterval of $\left[ x_{0},x\right] $.
However, (\ref{bc1}) implies $\frac{d}{dx}\int \limits_{\mathcal{Y}}dy$ $%
q\left( \left. y\right \vert x\right) =0$. Therefore, the Euler-Lagrange
equation in this example does not have a solution that would satisfy this
condition.

\section{Conclusion}

The rational inattention framework has gained popularity as an alternative
to the rational expectations approach to the decision under uncertainty. It
is based on a plausible assumption that an economic agent has a limited
amount of attention and allocates it optimally among available bits of
information. However, formalisation of the solution as the optimal choice of
conditional distribution of action given the exogenous distribution of the
external state is not a well-posed problem, and the solution, in general,
does not exist. This paper demonstrates that this approach may lead to a
solution in one special case of the linear-quadratic objective with the
Gaussian distribution of the state, under a specific choice of the model
parameters that has no obvious interpretation. It is unlikely to be
applicable to a wider set of problems that are of interest for economists.
Other solution concepts used in the rational inattention literature can
prove more fruitful in further developments.

\section*{References}

\begin{description}
\item Gabaix, X. (2019). \textquotedblleft Behavioural
inattention,\textquotedblright \ in: Douglas Bernheim, Stefano DellaVigna,
and David Laibson (Eds.), \textit{Handbook of Behavioural Economics},
Elsevier, forthcoming. (A 2017 draft is available as \textit{NBER Working
Paper} No. 24096.)

\item Sims, C.A. (2003). \textquotedblleft Implications of rational
inattention,\textquotedblright \  \textit{Journal of Monetary Economics} 50
(3), pp. 665--690.

\item Sims, C.A. (2005). \textquotedblleft Rational inattention:\ A research
agenda,\textquotedblright \  \textit{Deutsche Bundesbank Discussion Papers}
Series 1: Economic Studies, No. 34/2005.

\item Sims, C.A. (2006). \textquotedblleft Rational inattention: Beyond the
linear-quadratic case,\textquotedblright \  \textit{American Economic Review
Papers and Proceedings} 96(2), pp. 158--163.

\item Smirnov, B.I., Krylov, V.I., Kantorovich, L.V. (1933). \textit{%
Calculus of Variations}. Kubuch: Leningrad State University. (\textit{%
Variatsionnoe ischislenie}, in Russian. Available online at
http://books.e-heritage.ru/book/10073298)

\item Wiederholt, M. (2017) \textquotedblleft Rational
Inattention,\textquotedblright \ in: Steven N. Durlauf and Lawrence E. Blume
(Eds.),\  \textit{The New Palgrave Dictionary of Economics}. Online Edition.
\end{description}

\section*{Appendix}

\begin{description}
\item[Variational derivative] 
\end{description}

The variational derivative of functional%
\begin{equation*}
F\left[ f\right] \left( x\right) =\int \limits_{\mathcal{X}}dx\text{ }%
f\left( x\right) \varphi \left( x\right)
\end{equation*}%
of a scalar function of one variable, $f\left( x\right) $, is calculated as%
\begin{eqnarray*}
\left. \frac{\delta F\left[ f\left( x\right) \right] }{\delta f\left(
x\right) }\right \vert _{x=x_{0}} &=&\lim_{\varepsilon \rightarrow 0}\frac{1%
}{\varepsilon }\left \{ \int \limits_{\mathcal{X}}dx\text{ }\left[ f\left(
x\right) +\varepsilon \cdot \delta \left( x-x_{0}\right) \right] \varphi
\left( x\right) -\int \limits_{\mathcal{X}}dx\text{ }f\left( x\right)
\varphi \left( x\right) \right \} \\
&=&\lim_{\varepsilon \rightarrow 0}\frac{1}{\varepsilon }\int \limits_{%
\mathcal{X}}dx\text{ }\left[ \varepsilon \cdot \delta \left( x-x_{0}\right) %
\right] \varphi \left( x\right) =\varphi \left( x_{0}\right) .
\end{eqnarray*}%
where $\delta \left( x-x_{0}\right) $ is Dirac's delta function.\footnote{%
See, for example, Engel E., and Dreizler, R. M. \textit{Density Functional
Theory: An advanced course, }p. 409. Springer: Theoretical and Mathematical
Physics Series, 2011.} Similarly, for a functional of a scalar function of
two variables, $f\left( x,y\right) $, given by%
\begin{equation*}
F\left[ f\right] \left( x,y\right) =\iint \limits_{\mathcal{X\times Y}}dx%
\text{ }dy\text{ }f\left( x,y\right) \varphi \left( x,y\right)
\end{equation*}%
the variational derivative is calculated as%
\begin{eqnarray}
\left. \frac{\delta F\left[ f\left( x,y\right) \right] }{\delta f\left(
x,y\right) }\right \vert _{x=x_{0},y=y_{0}} &=&\lim_{\varepsilon \rightarrow
0}\frac{1}{\varepsilon }\left \{ \iint \limits_{\mathcal{X\times Y}}dx\text{ 
}dy\text{ }\left[ f\left( x,y\right) +\varepsilon \cdot \delta \left(
x-x_{0}\right) \delta \left( y-y_{0}\right) \right] \varphi \left(
x,y\right) \right.  \notag \\
&&\left. -\iint \limits_{\mathcal{X\times Y}}dx\text{ }dy\text{ }f\left(
x,y\right) \varphi \left( x,y\right) \right \}  \label{deriv} \\
&=&\lim_{\varepsilon \rightarrow 0}\frac{1}{\varepsilon }\int \limits_{%
\mathcal{X}}dx\text{ }\left[ \varepsilon \cdot \delta \left( x-x_{0}\right)
\delta \left( y-y_{0}\right) \right] \varphi \left( x,y\right) =\varphi
\left( x_{0},y_{0}\right) .  \notag
\end{eqnarray}%
We need to take the derivative of the Lagrangean 
\begin{equation*}
\mathcal{L=}E\left[ U\left( X,Y\right) \right] +\lambda \left[ \kappa \ln 2-J%
\right] ,
\end{equation*}%
where%
\begin{equation*}
E\left[ U\left( X,Y\right) \right] =\iint_{\mathcal{X\times Y}}dx\text{ }dy%
\text{ }q\left( \left. y\right \vert x\right) p\left( x\right) U\left(
x,y\right)
\end{equation*}%
and%
\begin{equation*}
J=\iint_{\mathcal{X\times Y}}dx\text{ }dy\text{ }q\left( \left. y\right
\vert x\right) p\left( x\right) \ln \frac{q\left( \left. y\right \vert
x\right) }{\int \nolimits_{\mathcal{X}}dx\text{ }q\left( \left. y\right
\vert x^{\prime }\right) p\left( x^{\prime }\right) }
\end{equation*}%
with respect to $q\left( \left. y\right \vert x\right) $:%
\begin{equation*}
\left. \frac{\delta \mathcal{L}}{\delta q\left( \left. y\right \vert
x\right) }\right \vert _{x=x_{0},y=y_{0}}=\left. \frac{\delta E\left[
U\left( X,Y\right) \right] }{\delta q\left( \left. y\right \vert x\right) }%
\right \vert _{x=x_{0},y=y_{0}}-\lambda \left. \frac{\delta J}{\delta
q\left( \left. y\right \vert x\right) }\right \vert _{x=x_{0},y=y_{0}}.
\end{equation*}%
For the first term, using (\ref{deriv}),%
\begin{equation*}
\left. \frac{\delta E\left[ U\left( X,Y\right) \right] }{\delta q\left(
\left. y\right \vert x\right) }\right \vert _{x=x_{0},y=y_{0}}=p\left(
x_{0}\right) U\left( x_{0},y_{0}\right) .
\end{equation*}%
In the second term rewrite $J$ as $J=J_{1}-J_{2}$, where%
\begin{eqnarray*}
J_{1} &=&\iint \limits_{\mathcal{X\times Y}}dx\text{ }dy\text{ }q\left(
\left. y\right \vert x\right) p\left( x\right) \ln q\left( \left. y\right
\vert x\right) , \\
J_{2} &=&\iint \limits_{\mathcal{X\times Y}}dx\text{ }dy\text{ }q\left(
\left. y\right \vert x\right) p\left( x\right) \ln \left( \int \limits_{%
\mathcal{X}}dx^{\prime }\text{ }q\left( \left. y\right \vert x^{\prime
}\right) p\left( x^{\prime }\right) \right) .
\end{eqnarray*}%
For $J_{1}$, (\ref{deriv}) gives%
\begin{eqnarray*}
\left. \frac{\delta J_{1}}{\delta q\left( \left. y\right \vert x\right) }%
\right \vert _{x=x_{0},y=y_{0}} &=&\lim_{\varepsilon \rightarrow 0}\frac{1}{%
\varepsilon }\left \{ \iint \limits_{\mathcal{X\times Y}}dx\text{ }dy\text{ }%
\left[ q\left( \left. y\right \vert x\right) +\varepsilon \cdot \delta
\left( x-x_{0}\right) \delta \left( y-y_{0}\right) \right] p\left( x\right)
\right. \\
&&\times \ln \left( q\left( \left. y\right \vert x\right) +\varepsilon \cdot
\delta \left( x-x_{0}\right) \delta \left( y-y_{0}\right) \right) \\
&&\left. -\iint \limits_{\mathcal{X\times Y}}dx\text{ }dy\text{ }q\left(
\left. y\right \vert x\right) p\left( x\right) \ln q\left( \left. y\right
\vert x\right) \right \} .
\end{eqnarray*}%
In the second line%
\begin{eqnarray*}
&&\ln \left( q\left( \left. y\right \vert x\right) +\varepsilon \cdot \delta
\left( x-x_{0}\right) \delta \left( y-y_{0}\right) \right) =\ln \left(
q\left( \left. y\right \vert x\right) \left[ 1+\frac{\varepsilon \cdot
\delta \left( x-x_{0}\right) \delta \left( y-y_{0}\right) }{q\left( \left.
y\right \vert x\right) }\right] \right) \\
&=&\ln q\left( \left. y\right \vert x\right) +\ln \left( 1+\frac{\varepsilon
\cdot \delta \left( x-x_{0}\right) \delta \left( y-y_{0}\right) }{q\left(
\left. y\right \vert x\right) }\right) =\ln q\left( \left. y\right \vert
x\right) +\frac{\varepsilon \cdot \delta \left( x-x_{0}\right) \delta \left(
y-y_{0}\right) }{q\left( \left. y\right \vert x\right) }+O\left( \varepsilon
^{2}\right) .
\end{eqnarray*}%
Thus,%
\begin{eqnarray*}
\left. \frac{\delta J_{1}}{\delta q\left( \left. y\right \vert x\right) }%
\right \vert _{x=x_{0},y=y_{0}} &=&\lim_{\varepsilon \rightarrow 0}\frac{1}{%
\varepsilon }\left \{ \iint \limits_{\mathcal{X\times Y}}dx\text{ }dy\text{ }%
\left[ q\left( \left. y\right \vert x\right) +\varepsilon \cdot \delta
\left( x-x_{0}\right) \delta \left( y-y_{0}\right) \right] p\left( x\right)
\right. \\
&&\times \left[ \ln q\left( \left. y\right \vert x\right) +\frac{\varepsilon
\cdot \delta \left( x-x_{0}\right) \delta \left( y-y_{0}\right) }{q\left(
\left. y\right \vert x\right) }+O\left( \varepsilon ^{2}\right) \right] \\
&&\left. -\iint \limits_{\mathcal{X\times Y}}dx\text{ }dy\text{ }q\left(
\left. y\right \vert x\right) p\left( x\right) \ln q\left( \left. y\right
\vert x\right) \right \} \\
&=&\lim_{\varepsilon \rightarrow 0}\frac{1}{\varepsilon }\left \{ \iint
\limits_{\mathcal{X\times Y}}dx\text{ }dy\text{ }\left[ \varepsilon \cdot
\delta \left( x-x_{0}\right) \delta \left( y-y_{0}\right) \right] p\left(
x\right) \ln q\left( \left. y\right \vert x\right) \right. \\
&&\left. +q\left( \left. y\right \vert x\right) p\left( x\right) \frac{%
\varepsilon \cdot \delta \left( x-x_{0}\right) \delta \left( y-y_{0}\right) 
}{q\left( \left. y\right \vert x\right) }+O\left( \varepsilon ^{2}\right)
\right \} \\
&=&p\left( x_{0}\right) \left[ \ln q\left( \left. y_{0}\right \vert
x_{0}\right) +1\right] .
\end{eqnarray*}%
Similarly, for $J_{2}$, (\ref{deriv}) gives%
\begin{eqnarray*}
\left. \frac{\delta J_{2}}{\delta q\left( \left. y\right \vert x\right) }%
\right \vert _{x=x_{0},y=y_{0}} &=&\lim_{\varepsilon \rightarrow 0}\frac{1}{%
\varepsilon }\left \{ \iint \limits_{\mathcal{X\times Y}}dx\text{ }dy\text{ }%
\left[ q\left( \left. y\right \vert x\right) +\varepsilon \cdot \delta
\left( x-x_{0}\right) \delta \left( y-y_{0}\right) \right] p\left( x\right)
\right. \\
&&\times \ln \left( \int \limits_{\mathcal{X}}dx^{\prime }\text{ }\left[
q\left( \left. y\right \vert x^{\prime }\right) +\varepsilon \cdot \delta
\left( x^{\prime }-x_{0}\right) \delta \left( y-y_{0}\right) \right] p\left(
x^{\prime }\right) \right) \\
&&\left. -\iint \limits_{\mathcal{X\times Y}}dx\text{ }dy\text{ }q\left(
\left. y\right \vert x\right) p\left( x\right) \ln \left( \int \limits_{%
\mathcal{X}}dx^{\prime }\text{ }q\left( \left. y\right \vert x^{\prime
}\right) p\left( x^{\prime }\right) \right) \right \} .
\end{eqnarray*}%
In the second line of the expression above,%
\begin{eqnarray*}
&&\ln \left( \int \limits_{\mathcal{X}}dx^{\prime }\text{ }\left[ q\left(
\left. y\right \vert x^{\prime }\right) +\varepsilon \cdot \delta \left(
x^{\prime }-x_{0}\right) \delta \left( y-y_{0}\right) \right] p\left(
x^{\prime }\right) \right) \\
&=&\ln \left( \int \limits_{\mathcal{X}}dx^{\prime }\text{ }q\left( \left.
y\right \vert x^{\prime }\right) p\left( x^{\prime }\right) +\int \limits_{%
\mathcal{X}}dx^{\prime \prime }\text{ }\varepsilon \cdot \delta \left(
x^{\prime \prime }-x_{0}\right) \delta \left( y^{\prime \prime
}-y_{0}\right) p\left( x^{\prime \prime }\right) \right) \\
&=&\ln \left( \left[ \int \limits_{\mathcal{X}}dx^{\prime }\text{ }q\left(
\left. y\right \vert x^{\prime }\right) p\left( x^{\prime }\right) \right] %
\left[ 1+\varepsilon \cdot \delta \left( y-y_{0}\right) \frac{\int \limits_{%
\mathcal{X}}dx^{\prime \prime }\text{ }\delta \left( x^{\prime \prime
}-x_{0}\right) p\left( x^{\prime \prime }\right) }{\int \limits_{\mathcal{X}%
}dx^{\prime }\text{ }q\left( \left. y\right \vert x^{\prime }\right) p\left(
x^{\prime }\right) }\right] \right) \\
&=&\ln \left( \int \limits_{\mathcal{X}}dx^{\prime }\text{ }q\left( \left.
y\right \vert x^{\prime }\right) p\left( x^{\prime }\right) \right) +\ln
\left( \left[ 1+\varepsilon \cdot \delta \left( y-y_{0}\right) \frac{p\left(
x_{0}\right) }{\int \limits_{\mathcal{X}}dx^{\prime }\text{ }q\left( \left.
y\right \vert x^{\prime }\right) p\left( x^{\prime }\right) }\right] \right)
\\
&=&\ln \left( \int \limits_{\mathcal{X}}dx^{\prime }\text{ }q\left( \left.
y\right \vert x^{\prime }\right) p\left( x^{\prime }\right) \right)
+\varepsilon \cdot \delta \left( y-y_{0}\right) \frac{p\left( x_{0}\right) }{%
\int \limits_{\mathcal{X}}dx^{\prime }\text{ }q\left( \left. y\right \vert
x^{\prime }\right) p\left( x^{\prime }\right) }+O\left( \varepsilon
^{2}\right) .
\end{eqnarray*}%
Upon substitution, the derivative simplifies as the following:%
\begin{eqnarray*}
\left. \frac{\delta I_{2}}{\delta q\left( \left. y\right \vert x\right) }%
\right \vert _{x=x_{0},y=y_{0}} &=&\lim_{\varepsilon \rightarrow 0}\frac{1}{%
\varepsilon }\left \{ \iint \limits_{\mathcal{X\times Y}}dx\text{ }dy\text{ }%
\left[ q\left( \left. y\right \vert x\right) +\varepsilon \cdot \delta
\left( x-x_{0}\right) \delta \left( y-y_{0}\right) \right] p\left( x\right)
\right. \\
&&\times \left[ \ln \left( \int \limits_{\mathcal{X}}dx^{\prime }\text{ }%
q\left( \left. y\right \vert x^{\prime }\right) p\left( x^{\prime }\right)
\right) +\varepsilon \cdot \delta \left( y-y_{0}\right) \frac{p\left(
x_{0}\right) }{\int \limits_{\mathcal{X}}dx^{\prime }\text{ }q\left( \left.
y\right \vert x^{\prime }\right) p\left( x^{\prime }\right) }+O\left(
\varepsilon ^{2}\right) \right] \\
&&\left. -\iint \limits_{\mathcal{X\times Y}}dx\text{ }dy\text{ }q\left(
\left. y\right \vert x\right) p\left( x\right) \ln \left( \int \limits_{%
\mathcal{X}}dx^{\prime }\text{ }q\left( \left. y\right \vert x^{\prime
}\right) p\left( x^{\prime }\right) \right) \right \} \\
&=&\lim_{\varepsilon \rightarrow 0}\frac{1}{\varepsilon }\left \{ \iint
\limits_{\mathcal{X\times Y}}dx\text{ }dy\text{ }\varepsilon \cdot \delta
\left( x-x_{0}\right) \delta \left( y-y_{0}\right) p\left( x\right) \ln
\left( \int \limits_{\mathcal{X}}dx^{\prime }\text{ }q\left( \left. y\right
\vert x^{\prime }\right) p\left( x^{\prime }\right) \right) \right. \\
&&\left. +\iint \limits_{\mathcal{X\times Y}}dx\text{ }dy\text{ }q\left(
\left. y\right \vert x\right) p\left( x\right) \varepsilon \cdot \delta
\left( y-y_{0}\right) \frac{p\left( x_{0}\right) }{\int \limits_{\mathcal{X}%
}dx^{\prime }\text{ }q\left( \left. y\right \vert x^{\prime }\right) p\left(
x^{\prime }\right) }+O\left( \varepsilon ^{2}\right) \right \} \\
&=&p\left( x_{0}\right) \ln \left( \int \limits_{\mathcal{X}}dx^{\prime }%
\text{ }q\left( \left. y_{0}\right \vert x^{\prime }\right) p\left(
x^{\prime }\right) \right) +\frac{p\left( x_{0}\right) \int \limits_{%
\mathcal{X}}dx\text{ }q\left( \left. y_{0}\right \vert x\right) p\left(
x\right) }{\int \limits_{\mathcal{X}}dx^{\prime }\text{ }q\left( \left.
y\right \vert x^{\prime }\right) p\left( x^{\prime }\right) } \\
&=&p\left( x_{0}\right) \left[ \ln \left( g\left( y_{0}\right) \right) +1%
\right] ,
\end{eqnarray*}%
where in the last line 
\begin{equation*}
g\left( y\right) =\int \limits_{\mathcal{X}}dx\text{ }f\left( x,y\right)
=\int \limits_{\mathcal{X}}dx\text{ }q\left( \left. y\right \vert x\right)
p\left( x\right)
\end{equation*}%
is the marginal density. Putting $J_{1}$ and $J_{2}$ together gives%
\begin{eqnarray*}
\left. \frac{\delta J}{\delta q\left( \left. y\right \vert x\right) }\right
\vert _{x=x_{0},y=y_{0}} &=&p\left( x_{0}\right) \left[ \ln q\left( \left.
y_{0}\right \vert x_{0}\right) +1\right] -p\left( x_{0}\right) \left[ \ln
\left( g\left( y_{0}\right) \right) +1\right] \\
&=&p\left( x_{0}\right) \ln \frac{q\left( \left. y_{0}\right \vert
x_{0}\right) }{g\left( y_{0}\right) }=p\left( x_{0}\right) \ln \frac{q\left(
\left. y_{0}\right \vert x_{0}\right) p\left( x_{0}\right) }{g\left(
y_{0}\right) p\left( x_{0}\right) } \\
&=&p\left( x_{0}\right) \ln \frac{h\left( \left. x_{0}\right \vert
y_{0}\right) }{p\left( x_{0}\right) }
\end{eqnarray*}%
where 
\begin{equation*}
h\left( \left. x\right \vert y\right) =\frac{f\left( x,y\right) }{g\left(
y\right) }=\frac{q\left( \left. y\right \vert x\right) p\left( x\right) }{%
g\left( y\right) }
\end{equation*}%
is the marginal density.

Finally, 
\begin{equation*}
\left. \frac{\delta \mathcal{L}}{\delta q\left( \left. y\right \vert
x\right) }\right \vert _{x=x_{0},y=y_{0}}=p\left( x_{0}\right) \left[
U\left( x_{0},y_{0}\right) -\lambda \ln \frac{h\left( \left. x_{0}\right
\vert y_{0}\right) }{p\left( x_{0}\right) }\right] .
\end{equation*}

This differs from the result in Sims (2005, 2006), which was derived from
the Lagrangean defined as 
\begin{eqnarray*}
\mathcal{L} &\mathcal{=}&\iint_{\mathcal{X\times Y}}dx\text{ }dy\text{ }%
q\left( \left. y\right \vert x\right) p\left( x\right) U\left( x,y\right)
+\lambda \left[ \kappa -\iint_{\mathcal{X\times Y}}dx\text{ }dy\text{ }%
q\left( \left. y\right \vert x\right) p\left( x\right) \log \frac{q\left(
\left. y\right \vert x\right) }{\int \nolimits_{\mathcal{X}}dx\text{ }%
q\left( \left. y\right \vert x\right) p\left( x\right) }\right] \\
&&+\mu \left( x\right) \left( \int \limits_{\mathcal{Y}}dy\text{ }f\left(
x,y\right) -p\left( x\right) \right)
\end{eqnarray*}%
where $\mu \left( x\right) $ is the Lagrange multiplier (see equation (12)
in Sims, 2005, p. 12, with $c=y$, $w=x$). This expression does not appear to
be correct because in the right-hand side there is a sum of a functional,
which is a definite integral, and a function of a variable. The
Euler-Lagrange equation (see equation (5) in Sims, 2005, 2006) derived from
this expression, is, therefore, incorrect. However, formally, the solution
for $h\left( \left. x\right \vert y\right) $ used in Sims (2003, 2006)
resembles (\ref{soln}), -- with $p\left( x\right) $ replaced by $\mu \left(
x\right) $, -- which has led Sims to a conjecture that the solution is
invariant to $p\left( x\right) $ `\textit{as long as the density has full
support}' (Sims, 2006, p. 162). One can see that, on the contrary, the
solution crucially depends on $p\left( x\right) $ and may exist only in
exceptional cases for a specific choice of $p\left( x\right) $.

\begin{description}
\item[Conditional Gaussian distribution for the linear-quadratic loss case] 
\end{description}

Consider the problem of minimisation of the expected value of a
linear-quadratic loss function,%
\begin{equation*}
U\left( X,Y\right) =-\theta ^{2}Y^{2}+2\varphi YX-X^{2}+2bX+2cY,\text{ }%
\mathcal{X\times Y=R\times R}.
\end{equation*}%
This is a generalisation of the quadratic loss function ($\varphi =\theta =1$%
, $b=c=0$) considered in Sims (2003), where it is stated that `\textit{when
the }$X$\textit{\ distribution is Gaussian, it is not too hard to show that
the optimal form for }$q$\textit{\ is also Gaussian, so that }$Y$\textit{\
and }$X$\textit{\  \ end up jointly normaly distributed}' (p. 670). As I show
below, Gaussian $q$ as a solution of (\ref{main}) given Gaussian $p$ only
exists and satisfies the properties of a distribution function under certain
restrictions on all but one of the loss function parameters.

Let $X\sim N\left( \mu _{x},\sigma _{x}^{2}\right) $. With $N\left( \mu
_{\left. x\right \vert y},\sigma _{\left. x\right \vert y}^{2}\right) $ as a
guess for $h\left( \left. x\right \vert y\right) $, (\ref{soln}) implies:%
\begin{equation}
U\left( x,y\right) =\lambda \ln \frac{h\left( \left. x\right \vert y\right) 
}{p\left( x\right) },  \label{Uxy}
\end{equation}%
where%
\begin{eqnarray*}
p\left( x\right) &=&\frac{1}{\sqrt{2\pi \sigma _{x}^{2}}}\exp \left( -\frac{%
\left[ x-\mu _{x}\right] ^{2}}{2\sigma _{x}^{2}}\right) , \\
h\left( \left. x\right \vert y\right) &=&\frac{1}{\sqrt{2\pi \sigma _{\left.
x\right \vert y}^{2}}}\exp \left( -\frac{\left[ x-\mu _{\left. x\right \vert
y}\right] ^{2}}{2\sigma _{\left. x\right \vert y}^{2}}\right) \\
&=&\frac{1}{\sqrt{2\pi \sigma _{x}^{2}\left( 1-\rho ^{2}\right) }}\exp
\left( -\frac{\left[ x-\left( \mu _{x}+\rho \frac{\sigma _{x}}{\sigma _{y}}%
\left( y-\mu _{y}\right) \right) \right] ^{2}}{2\sigma _{x}^{2}\left( 1-\rho
^{2}\right) }\right) .
\end{eqnarray*}%
Then%
\begin{eqnarray*}
\ln \frac{h\left( \left. x\right \vert y\right) }{p\left( x\right) } &=&\ln 
\frac{\frac{1}{\sqrt{2\pi \sigma _{x}^{2}\left( 1-\rho ^{2}\right) }}\exp
\left( -\frac{\left[ x-\left( \mu _{x}+\rho \frac{\sigma _{x}}{\sigma _{y}}%
\left( y-\mu _{y}\right) \right) \right] ^{2}}{2\sigma _{x}^{2}\left( 1-\rho
^{2}\right) }\right) }{\frac{1}{\sqrt{2\pi \sigma _{x}^{2}}}\exp \left( -%
\frac{\left[ x-\mu _{x}\right] ^{2}}{2\sigma _{x}^{2}}\right) } \\
&=&\ln \frac{1}{\sqrt{\left( 1-\rho ^{2}\right) }}-\frac{1}{2\sigma
_{x}^{2}\left( 1-\rho ^{2}\right) }\left( \left[ x-\left( \mu _{x}+\rho 
\frac{\sigma _{x}}{\sigma _{y}}\left( y-\mu _{y}\right) \right) \right]
^{2}-\left( 1-\rho ^{2}\right) \left[ x-\mu _{x}\right] ^{2}\right)
\end{eqnarray*}%
In the last term,

\begin{eqnarray*}
&&\left[ x-\left( \mu _{x}+\rho \frac{\sigma _{x}}{\sigma _{y}}\left( y-\mu
_{y}\right) \right) \right] ^{2}-\left( 1-\rho ^{2}\right) \left[ x-\mu _{x}%
\right] ^{2} \\
&=&\rho ^{2}x^{2}-2\rho \frac{\sigma _{x}}{\sigma _{y}}xy+\rho ^{2}\frac{%
\sigma _{x}^{2}}{\sigma _{y}^{2}}y^{2} \\
&&-2x\left( \rho ^{2}\mu _{x}-\rho \frac{\sigma _{x}}{\sigma _{y}}\mu
_{y}\right) +2y\rho \frac{\sigma _{x}}{\sigma _{y}}\left( \mu _{x}-\rho 
\frac{\sigma _{x}}{\sigma _{y}}\mu _{y}\right)  \\
&&+\rho ^{2}\mu _{x}^{2}-2\rho \frac{\sigma _{x}}{\sigma _{y}}\mu _{x}\mu
_{y}+\rho ^{2}\frac{\sigma _{x}^{2}}{\sigma _{y}^{2}}\mu _{y}^{2}.
\end{eqnarray*}%
Upon substitution in (\ref{Uxy}),%
\begin{eqnarray*}
&&-\theta y^{2}+2\varphi yx-x^{2}+2bx+2cy \\
&=&\lambda \ln \frac{1}{\sqrt{\left( 1-\rho ^{2}\right) }}-\frac{\lambda }{%
2\sigma _{x}^{2}\left( 1-\rho ^{2}\right) }\times  \\
&&\left[ \rho ^{2}x^{2}-2\rho \frac{\sigma _{x}}{\sigma _{y}}xy+\rho ^{2}%
\frac{\sigma _{x}^{2}}{\sigma _{y}^{2}}y^{2}\right.  \\
&&-2x\left( \rho ^{2}\mu _{x}-\rho \frac{\sigma _{x}}{\sigma _{y}}\mu
_{y}\right) +2y\rho \frac{\sigma _{x}}{\sigma _{y}}\left( \mu _{x}-\rho 
\frac{\sigma _{x}}{\sigma _{y}}\mu _{y}\right)  \\
&&\left. +\rho ^{2}\mu _{x}^{2}-2\rho \frac{\sigma _{x}}{\sigma _{y}}\mu
_{x}\mu _{y}+\rho ^{2}\frac{\sigma _{x}^{2}}{\sigma _{y}^{2}}\mu _{y}^{2}%
\right] 
\end{eqnarray*}%
and equating the coefficients at the powers and the cross-product of $x$ and 
$y$, we obtain%
\begin{eqnarray*}
0 &=&\lambda \ln \frac{1}{\sqrt{\left( 1-\rho ^{2}\right) }}-\frac{\lambda }{%
2\sigma _{x}^{2}\left( 1-\rho ^{2}\right) }\left( \rho ^{2}\mu
_{x}^{2}-2\rho \frac{\sigma _{x}}{\sigma _{y}}\mu _{x}\mu _{y}+\rho ^{2}%
\frac{\sigma _{x}^{2}}{\sigma _{y}^{2}}\mu _{y}^{2}\right) , \\
\theta ^{2} &=&\frac{\lambda }{2\sigma _{x}^{2}\left( 1-\rho ^{2}\right) }%
\rho ^{2}\frac{\sigma _{x}^{2}}{\sigma _{y}^{2}}, \\
\varphi  &=&\frac{\lambda }{2\sigma _{x}^{2}\left( 1-\rho ^{2}\right) }\rho 
\frac{\sigma _{x}}{\sigma _{y}}, \\
1 &=&\frac{\lambda }{2\sigma _{x}^{2}\left( 1-\rho ^{2}\right) }\rho ^{2}, \\
b &=&\frac{\lambda }{2\sigma _{x}^{2}\left( 1-\rho ^{2}\right) }\left( \mu
_{x}-\rho \frac{\sigma _{x}}{\sigma _{y}}\mu _{y}\right) , \\
c &=&-\frac{\lambda }{2\sigma _{x}^{2}\left( 1-\rho ^{2}\right) }\rho \frac{%
\sigma _{x}}{\sigma _{y}}\left( \mu _{x}-\rho \frac{\sigma _{x}}{\sigma _{y}}%
\mu _{y}\right) ,
\end{eqnarray*}%
which simplifies to%
\begin{eqnarray}
0 &=&\frac{1}{2}\ln \frac{1}{1-\rho ^{2}}-\left( \mu _{x}^{2}-2\varphi \mu
_{x}\mu _{y}+\frac{\sigma _{x}^{2}}{\sigma _{y}^{2}}\mu _{y}^{2}\right) 
\label{free} \\
\theta ^{2} &=&\frac{\sigma _{x}^{2}}{\sigma _{y}^{2}}  \label{y2} \\
\varphi  &=&\frac{1}{\rho }\frac{\sigma _{x}}{\sigma _{y}}  \label{xy} \\
\frac{1}{\rho ^{2}} &=&\frac{\lambda }{2\sigma _{x}^{2}\left( 1-\rho
^{2}\right) }  \label{x2} \\
b &=&\left( \mu _{x}-\varphi \mu _{y}\right)   \label{x} \\
c &=&-\varphi \left( \mu _{x}-\rho ^{2}\varphi \mu _{y}\right)   \label{y}
\end{eqnarray}

When the information constraint is binding, $I=\kappa $, and so%
\begin{equation*}
\frac{1}{2}\ln \frac{1}{1-\rho ^{2}}=\frac{1}{2}\log _{2}\frac{1}{1-\rho ^{2}%
}\ln 2=\kappa \ln 2,
\end{equation*}%
which gives%
\begin{equation*}
\rho ^{2}=1-2^{-2\kappa }.
\end{equation*}%
Using this in (\ref{x2}) gives for the Lagrange multiplier%
\begin{equation*}
\widetilde{\lambda }=\lambda \ln 2=\frac{2\ln 2}{2^{2\kappa }-1}\sigma
_{x}^{2}.
\end{equation*}

The optimal conditional distribution $q\left( \left. y\right \vert x\right) $
is Gaussian with 
\begin{eqnarray*}
\mu _{\left. y\right \vert x} &=&\mu _{y}+\rho \frac{\sigma _{y}}{\sigma _{x}%
}\left( x-\mu _{x}\right) , \\
\sigma _{\left. y\right \vert x}^{2} &=&\sigma _{y}^{2}\left( 1-\rho
^{2}\right) ,
\end{eqnarray*}%
where $\mu _{y}$ and $\sigma _{y}^{2}$ are the mean and the variance of the
(Gaussian) marginal distribution of $Y$, $g\left( y\right) $, and are
obtained from (\ref{free})-(\ref{y}).

Observe that (\ref{free}) gives%
\begin{equation*}
\mu _{x}^{2}-2\varphi \mu _{x}\mu _{y}+\frac{\sigma _{x}^{2}}{\sigma _{y}^{2}%
}\mu _{y}^{2}=\kappa \ln 2.
\end{equation*}%
That is, when $\mu _{x}=0$ it must be the case that 
\begin{equation*}
\mu _{y}=\frac{\sigma _{y}}{\sigma _{x}}\sqrt{\kappa \ln 2}\neq 0.
\end{equation*}%
This is contrary to the example in Gabaix (2019), who states that in Sims's
framework with the quadratic loss function\footnote{%
In Sims (2003) the loss function is $U\left( X,Y\right) =-\left( Y-X\right)
^{2}$.}, $U\left( X,Y\right) =-\frac{1}{2}\left( Y-X\right) ^{2\text{ }}$and 
$X\sim N\left( 0,\sigma ^{2}\right) $ the optimal action\footnote{%
The action in Gabaix (2019) is denoted by $a$.} is $Y\sim N\left( 0,\rho
^{2}\sigma ^{2}\right) $. Gabaix (2019) re-states the optimisation problem
as in Sims (2003), asserts that the optimal action is given by $Y=mS$, and
shows that $m=\rho ^{2}$ with $\rho ^{2}=1-e^{-2\kappa }$ (using natural
logarithms in the definition of entropy). Here $S=X+\varepsilon $ is a noisy
signal received by the agent who does not observe the true realisation of $X$%
, and $\varepsilon \sim N\left( 0,\sigma _{\varepsilon }^{2}\right) $ is
independent of $X$. However, Gabaix does not show how he derived the optimal
action from the constrained optimisation of the functional, and so it is not
clear how the solution for $q\left( \left. y\right \vert x\right) $ gives $%
Y=mS$.

Moreover, (\ref{free})-(\ref{y}) restrict the admissible parameters in the
loss function, so that given $\kappa $ and $\left \{ \mu _{x},\sigma
_{x}^{2}\right \} $ only one out of four parameters is free. In particular,
for $\mu _{x}=0$ and $\theta =1$ we have $\mu _{y}=\sqrt{\kappa \ln 2}$ and 
\begin{eqnarray*}
\varphi &=&\frac{1}{\sqrt{1-2^{-2\kappa }}}, \\
b &=&-\sqrt{\frac{\kappa \ln 2}{1-2^{-2\kappa }}}, \\
c &=&\sqrt{\kappa \ln 2},
\end{eqnarray*}%
so that%
\begin{equation*}
U\left( x,y\right) =-y^{2}+\frac{2}{\sqrt{1-2^{-2\kappa }}}yx-x^{2}-2\sqrt{%
\frac{\kappa \ln 2}{1-2^{-2\kappa }}}x+2\sqrt{\kappa \ln 2}y.
\end{equation*}

\end{document}